\documentclass[prd,amsmath,amssymb,longbibliography,superscriptaddress,twocolumn,nofootinbib,10pt]{revtex4-1}

\pdfoutput=1

\usepackage{graphicx}
\usepackage{dcolumn}
\usepackage{bm}
\usepackage{amssymb}
\usepackage{latexsym}
\usepackage{booktabs}
\usepackage{amsmath}
\usepackage{multirow}
\usepackage{enumitem}

\usepackage[colorlinks=true, linkcolor=blue, citecolor=blue]{hyperref}

\begin{document}

\title{Comparison of dark energy models using late-universe observations}

\author{Peng-Ju Wu}\email{wupengju@nxu.edu.cn}
\affiliation{School of Physics, Ningxia University, Yinchuan 750021, China}

\begin{abstract}
In the framework of general relativity, dark energy was proposed to explain the cosmic acceleration. A pivotal inquiry in cosmology is to determine whether dark energy is the cosmological constant, and if not, the challenge lies in constraining how it evolves with time. In this paper, we utilize the latest observational data to constrain some typical dark energy models, and make a comparison for them according to their capabilities of fitting the current data. Our study is confined to late-universe observations, including the baryon acoustic oscillation, type Ia supernova, cosmic chronometer, and strong gravitational lensing time delay data. We employ the Akaike information criterion (AIC), deviance information criterion (DIC),  and Bayesian information criterion (BIC) to assess the worth of models. The AIC and DIC analyses indicate that all dark energy models outperform the $\Lambda$CDM model. However, the BIC analysis leaves room for $\Lambda$CDM due to its heavier penalty on the model complexity. Compared to $\Lambda$CDM, most dark energy models are robustly supported by AIC and DIC while being explicitly disfavored by BIC. The models that are robustly favored by AIC and DIC and not explicitly disfavored by BIC include the $w$CDM, interacting dark energy, and Ricci dark energy models. Furthermore, we observe that an alternative modified gravity model exhibits superior performance when compared with $\Lambda$CDM across all information criteria.
\end{abstract}

\maketitle
\section{Introduction}\label{sec1}
It is generally believed that cosmic acceleration originates from the so-called dark energy, which has negative pressure and accounts for approximately 70\% of the energy budget of the universe. The investigation into the nature of dark energy has become one of the most important issues in the field of fundamental physics. In the standard model of cosmology, dark energy is considered as the cosmological constant $\Lambda$, which is associated to the zero-point vacuum energy density of quantum fields. The $\Lambda$ cold dark matter ($\Lambda$CDM) model has shown excellent agreement with various observations over an extended period \cite{Aghanim:2018eyx}. However, it faces some theoretical challenges \cite{Sahni:1999gb,Park:2017xbl}. Recently, it has also encountered several measurement inconsistencies, particularly regarding the Hubble constant\footnote{It should be noted that the Hubble tension primarily arises between Cepheid-SNIa distance ladders and CMB analysis; however, many other low-redshift $H_0$ measurements are consistent with the CMB value, see Refs.~\cite{Dominguez:2019jqc, Park:2019emi, Lin:2019zdn, Freedman:2020dne, Birrer:2020tax,Efstathiou:2020wxn,Boruah:2020fhl,Freedman:2021ahq,Wu:2021jyk,Cao:2022ugh,Chen:2024gnu,Freedman:2024eph,TDCOSMO:2025dmr}.} $H_0$, cosmic curvature parameter\footnote{There are also some studies argue against the existence of a curvature tension, see Refs.~\cite{Efstathiou:2020wem,Liu:2024yib, deCruzPerez:2024shj,Fortunato:2025qxc}.} $\Omega_K$, and the amplitude of the matter power spectrum\footnote{There is probably no longer a $\sigma_8$ measurement inconsistency, at least the main one between weak lensing and CMB observations has been resolved, see Refs.~\cite{Wright:2025xka,Stolzner:2025htz,SPT-3G:2025zuh}.} $\sigma_8$ \cite{Verde:2019ivm,Riess:2021jrx,Vagnozzi:2023nrq,Jiang:2024xnu,Vagnozzi:2020rcz,Vagnozzi:2020dfn,DESI:2024mwx,Wu:2024faw}. All of these issues have driven researchers to seek underestimated systematic errors or new physics beyond the standard model.

Cosmologists have conceived many physically-rich dark energy models to address the difficulties encountered by the standard model. For instance, a slowly rolling scalar field can also generate negative pressure, thereby driving the current phase of cosmic acceleration. Such light scalar fields provide a potential mechanism for dynamical dark energy \cite{Peebles:1987ek, Ratra:1987rm, Zlatev:1998tr,Steinhardt:1999nw,Wolf:2024stt,Payeur:2024dnq,Wolf:2024eph,Ye:2024ywg}. Generally, one can characterize the properties of dark energy through parametrizing its equation of state (EoS) as $w=p_{\rm de}/\rho_{\rm de}$, where $p_{\rm de}$ and $\rho_{\rm de}$ represent the pressure and energy density of dark energy, respectively. The simplest parametrized model, where $w$ is a constant, is commonly referred to as the $w$CDM model \cite{Escamilla:2023oce}. This model subsumes two distinct scenarios of dark energy: Quintessence, which corresponds to cases where $w>-1$ \cite{Payeur:2024dnq}, and Phantom, which pertains to instances where $w<-1$. A more physical and realistic scenario is that $w$ varies with cosmic time, and one of the most common parametrizations is $w=w_0+w_a(1-a)$ \cite{Chevallier:2000qy,Linder:2002et}, where $a$ is the cosmic scale factor. For more dark energy EoS parametrizations, see Refs.~\cite{Wetterich:2004pv,Linder:2006sv,Jassal:2004ej,Ma:2011nc,Bamba:2012cp}.

A variety of extensions to the $\Lambda$CDM model and scalar field theories are worth investigating, such as taking into account the spatial geometry of the universe or the interaction between dark energy and dark matter. On the one hand, some studies indicate that our universe may not be spatially flat. For example, the cosmic microwave background (CMB) observations favor a closed universe \cite{Handley:2019tkm,DiValentino:2019qzk}, while the combination of non-CMB probes supports an open universe \cite{Wu:2024faw}. While the two results exhibit a statistical discrepancy, both imply a non-flat universe. It should be stressed that the dark energy EoS and $\Omega_K$ are degenerate. Thus, it is necessary to incorporate $\Omega_K$ into parameter space and examine its impact on the dark energy EoS measurement. On the other hand, there may be an interaction between dark energy and dark matter. From the view of particle physics, interactions are ubiquitous. Cosmologists suggest that dark energy may have a non-gravitational interaction with dark matter, which is described by the interacting dark energy (IDE) models \cite{Valiviita:2008iv,Koyama:2009gd,Clemson:2011an,Li:2014eha, Li:2013bya,Xia:2016vnp}. According to some recent studies, there is evidence suggesting the presence of interaction between dark sectors; however, the conclusion is contingent upon the underlying IDE models \cite{DiValentino:2019ffd,DiValentino:2019jae,Giare:2024smz,Li:2024qso,Silva:2025hxw,Pan:2025qwy,Li:2025owk}.

Some dark energy models are built upon deep theoretical considerations. For example, the holographic dark energy (HDE) model is posited to have a quantum gravity origin. It is constructed by incorporating gravitational effects into the effective quantum field theory via the holographic principle, which proposes that the number of degrees of freedom in a spatial region is finite to prevent the black hole formation \cite{Cohen:1998zx,Li:2004rb}. The model not only presents a unique perspective on the nature of dark energy but also demonstrates the potential to resolve the puzzles faced by the standard model. Its theoretical variant, the Ricci dark energy (RDE) model \cite{Gao:2007ep}, has also garnered a lot of attention. The distinction between HDE and RDE hinges on the selection of the infrared cutoff utilized in the computation of dark energy density. Specifically, HDE opts for the radius of the future event horizon of the universe as the infrared cutoff, while RDE employs the average radius of the Ricci scalar curvature as the infrared cutoff. Investigating these two models is crucial for exploring quantum gravity-based explanations of dark energy \cite{Li:2024qus,Li:2024hrv,Tang:2024gtq,Trivedi:2024dju,Moradpour:2020dfm,Kibe:2021gtw}.

In physics, there are actually two mechanisms that can account for the accelerated expansion of the universe, i.e., assuming the existence of dark energy or modifying the gravity on large scales. The latter is termed modified gravity (MG) theory, which can produce ``effective dark energy'' scenarios mimicking the behavior of actual dark energy. By setting aside the issues of structure growth, it is possible to consider such models as alternatives to dark energy models. A prominent exemplar of such theoretical constructs is the Dvali-Gabadze-Porrati (DGP) model \cite{Dvali:2000hr}. This model originates within the framework of braneworld theories, which propose that our observable universe is a brane embedded in a higher-dimensional space, known as the ``bulk''. In the DGP model, gravity exhibits a tendency to leak into the bulk at cosmological scales, which contributes to the observed acceleration of our universe. The theoretical variant of DGP, the $\alpha$DM model \cite{Dvali:2003rk}, has also been widely studied due to its ability to better fit the historical observational data.

Facing numerous rival dark energy models, the primary objective is to determine which one represents the true dark energy scenario. However, achieving this is exceedingly challenging. A more practical goal is to discern which ones can better fit the observational data. \cite{Xu:2016grp}. Currently, there exist some measurement discrepancies between the early- and late-universe observations. For instance, the $H_0$ value measured by the Cepheid-supernova distance ladder is in above $5\sigma$ tension with that derived from the CMB observations assuming $\Lambda$CDM \cite{Riess:2021jrx}. When integrating observations to constrain cosmological parameters, it is necessary to consider the potential measurement inconsistencies among different datasets. The existence of these tensions diminish the justifiability of combining CMB with late-universe probes for parameter inference\footnote{It should be stressed that although discrepancies exist between early- and late-time data, this does not imply that all observations between early- and late universe are inconsistent. Additionally, inconsistencies may also exist among the various late-universe cosmological probes themselves.} The motivation of this work is to constrain various dark energy models using the combination of late-universe observations and assess the worth of them.

We consider four non-CMB probes, including the baryon acoustic oscillation (BAO, standard ruler), type Ia supernova (SN, standard candle), cosmic chronometer (CC, standard clock), and strong gravitational lensing time delay (TD) observations. For an extended period, these non-CMB observations failed to effectively constrain dark energy models. However, recent advancements may have altered this situation. Recently, the DESI collaboration released the high-precision BAO data \cite{DESI:2024uvr,DESI:2024lzq,DESI:2024mwx,DESI:2025fii}, the DES program published the high-quality samples of SN Ia \cite{DES:2024tys}, and to date, there are 32 CC data \cite{Moresco:2022phi} and 7 TD data \cite{Suyu:2009by,Jee:2019hah,Suyu:2013kha,Chen:2019ejq,Wong:2016dpo,Birrer:2018vtm,Rusu:2019xrq,DES:2019fny,Agnello:2017mwu} have become available for parameter estimation. We shall employ these late-universe observational data to constrain dark energy models.

The paper is organized as follows: Sec.\,\ref{sec2} outlines the methodology, Sec.\,\ref{sec3} details the observational data and their processing procedures, Sec.\,\ref{sec4} compares different dark energy models, and Sec.\,\ref{sec5} concludes with key findings and their cosmological implications.

\section{methodology}\label{sec2}
The constraints on cosmological parameters stem from precise measurements of cosmological distances; we start with a review of the definitions and distinctions between key distance metrics in cosmology. Firstly, the comoving distance is defined as
\begin{equation}\label{DC}
D_{\rm C}(z)=\int_{0}^{z} \frac{c{\rm d}z'}{H(z')}=\frac{c}{H_0}\int_{0}^{z} \frac{{\rm d}z'}{E(z')},
\end{equation}
where $E(z)=H(z)/H_0$ is the reduced Hubble parameter. The transverse comoving distance is given by \cite{Hogg:1999ad}
\begin{equation}\label{DM}
D_{\rm M}(z)=\left\{
\begin{aligned}
&\frac{c}{H_0\sqrt{\Omega_K}}\sinh{\left[\frac{H_0\sqrt{\Omega_K}}{c}D_{\rm C}(z)\right]} & {\rm if}\ \Omega_{K}>0 ,  \\
&D_{\rm C}(z)                                                                             & {\rm if}\ \Omega_{K}=0 , \\
&\frac{c}{H_0\sqrt{-\Omega_K}}\sin{\left[\frac{H_0\sqrt{-\Omega_K}}{c}D_{\rm C}(z)\right]}  & {\rm if}\ \Omega_{K}<0 .
\end{aligned}
\right.
\end{equation}
Then the luminosity distance, angular diameter distance, and Hubble distance can be written as
\begin{align}\label{DLDADH}
D_{\rm L}=D_{\rm M}\,(1+z);\ D_{\rm A}=D_{\rm M}/(1+z);\  D_{\rm H}=c/H.
\end{align}
The measurement of these cosmological distances enables the imposition of constraints on parameters embedded in the $E(z)$ formalism, and the primary distinctions among different cosmological models are reflected in the form of $E(z)$.

In this work, we adopt the Markov Chain Monte Carlo (MCMC) analysis to infer the probability distributions of cosmological parameters using the observational data, by maximizing the likelihood $L\propto\rm{exp}(-\chi^2/2)$. The $\chi^2$ function of each dataset can be written as $\Delta\boldsymbol{D}^T\boldsymbol{C}^{-1}\Delta\boldsymbol{D}$, where $\Delta\boldsymbol{D}$ is the vector of observable residuals representing the difference between observation and theory, and $\boldsymbol{C}$ is the covariance matrix. We measure the convergence of the chains by checking that all parameters have $R-1<0.01$, where $R$ is the potential scale reduction factor of the Gelman-Rubin diagnostics. We shall make a comparison for different dark energy models according to their capabilities of fitting the current data.

As models become more complex (i.e., have more free parameters), they tend to fit the observational data better, which can lead to a lower $\chi^2$ value. Therefore, the $\chi^2$ comparison is unfair for comparing dark energy models. In this work, we adopt the Akaike information criterion (AIC), Bayesian information criterion (BIC), and deviance information criterion (DIC) to compare different models. AIC is defined as
\begin{align}
\text{AIC} =\chi^2_{\rm min}+ 2k,
\end{align}
where $k$ is the number of free parameters. A model with a smaller AIC is more favored by observations. BIC is calculated by
\begin{align}
\text{BIC} = \chi^2_{\rm min}+ k \ln N,
\end{align}
where $N$ is the number of data points used in the fitting.
DIC is calculated by
\begin{align}
\text{DIC} = \chi^2_{\rm min}+ 2k_{\rm eff},
\end{align}
where $k_{\rm eff}=\langle \chi^2\rangle -\chi^2_{\rm min} $ is the the number of effectively constrained parameters. Here, the angular brackets denote the average over the posterior distribution.

In practice, we are not concerned with the absolute values of A/B/DIC, but the relative value between different models, i.e., $\Delta{\rm A/B/DIC}$. In this work, the flat $\Lambda$CDM model is adopted as the reference. Negative $\Delta{\rm A/B/DIC}$ values denote superior fit to the observational data relative to flat $\Lambda$CDM, whereas positive values indicate inferior fit. Generally, a negative $\Delta{\rm A/B/DIC}$ value within the range $[-2, 0)$ indicates weak evidence in favor of the model, a value within $[-6, -2)$ indicates positive evidence in favor, a value within $[-10, -6)$ indicates strong evidence in favor, and a value less than $-10$ indicates very strong evidence in favor. Conversely, a $\Delta{\rm A/B/DIC}$ value within the range $(0, 2]$ indicates weak evidence against the model, a value within $(2, 6]$ indicates positive evidence against, a value within $(6, 10]$ indicates strong evidence against, and a value greater than 10 indicates very strong evidence against. The penalty terms of AIC and DIC depend exclusively on the number of parameters, not sample size. Therefore, once the number of data points is large, the results may favor the model with more parameters. To further penalize the model complexity, BIC takes into account the number of data points. Using AIC, BIC, and DIC together gives a complete picture when comparing models.

\section{Observational data}\label{sec3}

$\bullet$ {\bf Baryon Acoustic Oscillations.} We consider the recently released DESI BAO data and summarize them in Table~\ref{BAO}. The data points with same redshifts are correlated, and the covariance matrices can be found in the website.\footnote{\url{https://data.desi.lbl.gov/doc/releases/}}
\begin{table}
\renewcommand\arraystretch{1}
\caption{The BAO measurements from DESI. Here, $r_{\rm d}$ is the sound horizon at the drag epoch and $D_{\rm V}(z)$ is the comoving volume-averaged distance which is calculated by $\left[D_{\rm M}(z)\right]^{2/3}\left[cD_{\rm H}(z)\right]^{1/3}$.}
\label{BAO}
\centering
\begin{tabular}{l p{1.7cm}<{\centering} p{2.2cm}<{\centering} p{2cm}<{\centering}}
\bottomrule[1pt]
& Redshift $z$  &Observable                     & Value                     \\
\bottomrule[1pt]
& 0.30          & $D_{\rm V}/r_{\rm d}$         & $7.93\pm0.15$             \\
& 0.51          & $D_{\rm M}/r_{\rm d}$         & $13.62\pm0.25$            \\
& 0.51          & $D_{\rm H}/r_{\rm d}$         & $20.98\pm0.61$            \\
& 0.71          & $D_{\rm M}/r_{\rm d}$         & $0.497\pm0.045$           \\
& 0.71          & $D_{\rm H}/r_{\rm d}$         & $13.38\pm0.18$            \\
& 0.93          & $D_{\rm M}/r_{\rm d}$         & $22.43\pm0.48$            \\
& 0.93          & $D_{\rm H}/r_{\rm d}$         & $0.459\pm0.038$           \\
& 1.32          & $D_{\rm M}/r_{\rm d}$         & $17.65\pm0.30$            \\
& 1.32          & $D_{\rm H}/r_{\rm d}$         & $19.78\pm0.46$            \\
& 1.49          & $D_{\rm V}/r_{\rm d}$         & $0.473\pm0.041$           \\
& 2.33          & $D_{\rm M}/r_{\rm d}$         & $19.5\pm1.0$              \\
& 2.33          & $D_{\rm H}/r_{\rm d}$         & $19.6\pm2.1$              \\
\bottomrule[1pt]
\end{tabular}
\end{table}
Note that the sound horizon at the drag epoch is treated as a free parameter in the MCMC analysis, which can ensure that the derived constraints are independent of the early-universe observational priors.

$\bullet$ {\bf Type Ia Supernovae.} We consider the DES sample of 1829 distant SNe~Ia spanning $0.025<z<1.3$ \cite{DES:2024tys}. Compared to the PantheonPlus sample, the dataset quintuples the number of SNe beyond $z > 0.5$. The distance modulus for an SN~Ia is defined by $\mu=m_{B}-M_{B}$, where $m_{B}$ is the observed apparent magnitude in the rest-frame $B$-band, and $M_{B}$ denotes the absolute magnitude. The theoretical distance modulus is calculated by
\begin{align}
\mu_{\rm th}(z)=5 \log \left[\frac{D_{\rm L}(z)}{\mathrm{Mpc}}\right]+25.
\end{align}
The observational data, encompassing distance moduli, their associated uncertainties and covariance between distinct data points, are accessible on the website.\footnote{\url{https://github.com/des-science/DES-SN5YR}}

\begin{table}
\renewcommand\arraystretch{1}
\caption{The 32 $H(z)$ measurements obtained with the CC method.}
\label{CCHz}
\centering
\begin{tabular}{l p{1.8cm}<{\centering} p{2.4cm}<{\centering} p{1.8cm}<{\centering}}
\bottomrule[1pt]
& Redshift $z$ & $H(z)\ [\,\rm km/s/Mpc\,]$            & Reference                       \\
\bottomrule[1pt]
& 0.07         & 69  $\pm$      19.6                  & \citep{Zhang:2012mp}             \\
& 0.09         & 69    $\pm$   12                    & \citep{Simon:2004tf}             \\
& 0.12         & 68.6  $\pm$26.2                  & \citep{Zhang:2012mp}             \\
& 0.17         & 83   $\pm$ 8                     & \citep{Simon:2004tf}             \\
& 0.179        & 75  $\pm$ 4                     & \citep{Moresco:2012jh}           \\
& 0.199        & 75   $\pm$ 5                     & \citep{Moresco:2012jh}           \\
& 0.2          & 72.9  $\pm$ 29.6                  & \citep{Zhang:2012mp}             \\
& 0.27         & 77     $\pm$14                    & \citep{Simon:2004tf}             \\
& 0.28         & 88.8  $\pm$36.6                  & \citep{Zhang:2012mp}             \\
& 0.352        & 83    $\pm$ 14                    & \citep{Moresco:2012jh}           \\
& 0.38         & 83    $\pm$ 13.5                  & \citep{Moresco:2016mzx}          \\
& 0.4          & 95     $\pm$ 17                    & \citep{Simon:2004tf}             \\
& 0.4004       & 77    $\pm$ 10.2                  & \citep{Moresco:2016mzx}          \\
& 0.425        & 87.1  $\pm$ 11.2                  & \citep{Moresco:2016mzx}          \\
& 0.445        & 92.8  $\pm$ 12.9                  & \citep{Moresco:2016mzx}          \\
& 0.47         & 89   $\pm$ 49.6                  & \citep{Ratsimbazafy:2017vga}     \\
& 0.4783       & 80.9   $\pm$ 9                     & \citep{Moresco:2016mzx}          \\
& 0.48         & 97    $\pm$ 62                    & \citep{Stern:2009ep}             \\
& 0.593        & 104   $\pm$ 13                    & \citep{Moresco:2012jh}           \\
& 0.68         & 92    $\pm$ 8                     & \citep{Moresco:2012jh}           \\
& 0.75         & 98.8  $\pm$ 33.6                  & \citep{Borghi:2021rft}           \\
& 0.781        & 105   $\pm$ 12                    & \citep{Moresco:2012jh}           \\
& 0.875        & 125   $\pm$ 17                    & \citep{Moresco:2012jh}           \\
& 0.88         & 90    $\pm$ 40                    & \citep{Stern:2009ep}             \\
& 0.9          & 117    $\pm$ 23                    & \citep{Simon:2004tf}             \\
& 1.037        & 154   $\pm$ 20                    & \citep{Moresco:2012jh}           \\
& 1.3          & 168    $\pm$ 17                    & \citep{Simon:2004tf}             \\
& 1.363        & 160   $\pm$33.6                  & \citep{Moresco:2015cya}          \\
& 1.43         & 177    $\pm$ 18                    & \citep{Simon:2004tf}             \\
& 1.53         & 140   $\pm$ 14                    & \citep{Simon:2004tf}             \\
& 1.75         & 202    $\pm$ 40                    & \citep{Simon:2004tf}             \\
& 1.965        & 186.5  $\pm$ 50.4                  & \citep{Moresco:2015cya}          \\
\bottomrule[1pt]
\end{tabular}
\end{table}
$\bullet$ {\bf Cosmic Chronometers.} The CC method provides a direct way to measure the Hubble parameter. We summarize the latest 32 CC $H(z)$ measurements in Table~\ref{CCHz}. Among them, 15 data points from Refs.~\cite{Moresco:2012jh,Moresco:2016mzx,Moresco:2015cya} are correlated, and the covariance matrix can be found in the website\footnote{\url{https://gitlab.com/mmoresco/CCcovariance}}.

$\bullet$ {\bf Strong Gravitational Lensing.} We consider the existing seven SGL systems with time-delay distance and angular diameter distance measurements, and summarize them in Table~\ref{TD}. The time-delay distance is defined by
\begin{align}
\label{SGL-TD}
D_{\Delta t}\equiv(1+z_{\rm l})\displaystyle{\frac{D_{\rm l}D_{\rm s}}{D_{\rm ls}}},
\end{align}
where $D_{\rm l}$, $D_{\rm s}$, and $D_{\rm ls}$ are the angular diameter distances between observer and lens, between observer and source, and between lens and source, respectively. Note that the errors of data will be approximated to a Gaussian form. For more details, we refer readers to the website.\footnote{\url{https://zenodo.org/records/3633035}}

\begin{table}
\renewcommand\arraystretch{1.2}
\caption{The time-delay distances and angular diameter distances for TD measurements. Here, $z_{\rm l}$ and $z_{\rm s}$ are the redshifts of lens and source respectively.}
\label{TD}
\centering
\begin{tabular}{l  p{1.2cm}<{\centering} p{1.2cm}<{\centering} p{1.7cm}<{\centering} p{1.7cm}<{\centering} p{1.5cm}<{\centering}}
\bottomrule[1pt]
& $z_{\rm l}$   & $z_{\rm s}$   & $D_{\Delta t}\ [\rm Mpc]$     & $D_{\rm l}\ [\rm Mpc]$    & References                                      \\
\bottomrule[1pt]
& 0.6304        & 1.394         & $5156^{+296}_{-236}$          & $1228^{+177}_{-151}$      & \cite{Suyu:2009by,Jee:2019hah}                 \\
& 0.295         & 0.654         & $2096^{+98}_{-83}$            & $804^{+141}_{-112}$       & \cite{Suyu:2013kha,Chen:2019ejq}               \\
& 0.4546        & 1.693         & $2707^{+183}_{-168}$          & $-$                       & \cite{Wong:2016dpo,Chen:2019ejq}               \\
& 0.745         & 1.789         & $5769^{+589}_{-471}$          & $1805^{+555}_{-398}$      & \cite{Birrer:2018vtm}                          \\
& 0.6575        & 1.662         & $4784^{+399}_{-248}$          & $-$                       & \cite{Rusu:2019xrq}                            \\
& 0.311         & 1.722         & $1470^{+137}_{-127}$          & $697^{+186}_{-144}$       & \cite{Chen:2019ejq}                            \\
& 0.597         & 2.375         & $3382^{+146}_{-115}$          & $1711^{+376}_{-280}$      & \cite{DES:2019fny,Agnello:2017mwu}             \\
\bottomrule[1pt]
\end{tabular}
\end{table}

\section{dark energy models}\label{sec4}
In this section, we describe the dark energy models selected for analysis and report the constraint results from the combination of late-universe observations. We utilize the same datasets to constrain these models and conduct a comparison for them. For organizational purposes, we divide these models into six distinct groups:
\begin{enumerate}[label=(\alph*)]
    \item Cosmological constant model.
    \item Dark energy models with a parameterized EoS.
    \item Dark energy models assuming a non-flat universe.
    \item Interacting dark energy models.
    \item Holographic dark energy models.
    \item Effective dark energy models (MG theories).
\end{enumerate}
The constraint results for dark energy models are shown in Table~\ref{tab:results1}. The results of the model comparison using the information criteria are summarized in Table~\ref{tab:results2}. The posterior distribution contours are shown in Figs.~\ref{LCDM}--\ref{alphaDE}, with no statistically significant discrepancies among late-universe datasets observed.

\begin{table*}[!htb]
\caption{Fit results for the dark energy models by using the observational data. Here $H_0$ is in units of $\rm km/s/Mpc$.}
\label{tab:results1}
\setlength{\tabcolsep}{2mm}
\renewcommand{\arraystretch}{1.4}
\begin{center}{\centerline{
\begin{tabular}{lllllll}
\bottomrule[1pt]
Model   &Parameter                  &                                           &                                       &                                           \\
\bottomrule[1pt]
{$\Lambda$CDM}
        &$H_0=72.7^{+1.2}_{-1.2}$   &$\Omega_{\rm m}=0.317^{+0.011}_{-0.011}$   &                                       &                                           \\
{DGP}
        &$H_0=71.5^{+1.2}_{-1.2}$   &$\Omega_{\rm m}=0.245^{+0.010}_{-0.010}$   &                                       &                                           \\
{o$\Lambda$CDM}
        &$H_0=72.8^{+1.1}_{-1.3}$   &$\Omega_{\rm m}=0.286^{+0.019}_{-0.019}$   &$\Omega_{K}=0.106^{+0.056}_{-0.056}$   &                                           \\
{I$\Lambda$CDM}
        &$H_0=71.9^{+1.3}_{-1.3}$   &$\Omega_{\rm m}=0.380^{+0.027}_{-0.027}$   &$\beta=-0.36^{+0.14}_{-0.14}$          &                                           \\
{$\alpha$DE}
        &$H_0=72.1^{+1.3}_{-1.3}$   &$\Omega_{\rm m}=0.284^{+0.022}_{-0.017}$   &$\alpha=0.56^{+0.22}_{-0.22}$          &                                           \\
{HDE}
        &$H_0=71.8^{+1.3}_{-1.3}$   &$\Omega_{\rm m}=0.266^{+0.014}_{-0.014}$   &$c=1.112^{+0.093}_{-0.120}$            &                                           \\
{RDE}
        &$H_0=71.8^{+1.2}_{-1.2}$   &$\Omega_{\rm m}=0.212^{+0.012}_{-0.012}$   &$\gamma=0.554^{+0.019}_{-0.022}$       &                                           \\
{$w$CDM}
        &$H_0=71.9^{+1.2}_{-1.2}$   &$\Omega_{\rm m}=0.295^{+0.014}_{-0.014}$   &$w=-0.877^{+0.046}_{-0.046}$           &                                           \\
{I$w$CDM}
        &$H_0=71.9^{+1.2}_{-1.4}$   &$\Omega_{\rm m}=0.53^{+0.31}_{-0.14}$      &$w=-1.72^{+1.10}_{-0.45}$              &$\beta=-2.5^{+3.3}_{-1.3}$                 \\
{o$w$CDM}
        &$H_0=72.0^{+1.3}_{-1.3}$   &$\Omega_{\rm m}=0.289^{+0.019}_{-0.019}$   &$w=-0.890^{+0.065}_{-0.050}$           &$\Omega_{K}=0.023^{+0.073}_{-0.073}$       \\
{CPL}
        &$H_0=71.0^{+1.3}_{-1.3}$   &$\Omega_{\rm m}=0.329^{+0.018}_{-0.015}$   &$w_0=-0.725^{+0.084}_{-0.097}$         &$w_a=-1.32^{+0.65}_{-0.65}$                \\
{JBP}
        &$H_0=70.9^{+1.3}_{-1.3}$   &$\Omega_{\rm m}=0.324^{+0.017}_{-0.016}$   &$w_0=-0.64^{+0.11}_{-0.12}$            &$w_a=-2.23^{+1.1}_{-0.97}$                 \\
\bottomrule[1pt]
\end{tabular}}}
\end{center}
\end{table*}

\begin{table}[!htb]
\caption{Summary of the information criteria results.}
\label{tab:results2}
\setlength{\tabcolsep}{2mm}
\renewcommand{\arraystretch}{1.4}
\begin{center}{\centerline{
\begin{tabular}{llllll}
\bottomrule[1pt]
Model   & $\chi^2_{\rm min}$    &$\Delta\rm {AIC}$          &$\Delta\rm {BIC}$  &$\Delta\rm {DIC}$ \\
\bottomrule[1pt]
{$\Lambda$CDM}
        &1686.282                &0                             &0                      &0     \\
{DGP}
        &1684.453                &$-1.829$                  & $-1.829$          & $-1.798$  \\
{o$\Lambda$CDM}
        &1682.493                &$-1.789$                  &$3.752$            & $-1.814$   \\
{I$\Lambda$CDM}
        &1679.213               &$-5.069$                   &$0.472$            & $-4.710$   \\
{$\alpha$DE}
        &1680.915               &$-3.367$                   &$2.175$            & $-3.547$    \\
{HDE}
        &1680.787               &$-3.495$                   &$2.046$            & $-3.399$    \\
{RDE}
        &1679.183               &$-5.092$                   &$0.442$            & $-5.069$   \\
{$w$CDM}
        &1679.190               &$-5.092$                   &$0.449$            & $-4.972$  \\
{I$w$CDM}
        &1679.181               &$-3.101$                   &$7.982$            & $-4.644$  \\
{o$w$CDM}
        &1679.119              &$-3.163$                    &$7.920$            & $-2.873$  \\
{CPL}
        &1674.821               &$-7.461$                   &$3.622$            & $-7.389$ \\
{JBP}
        &1674.480               &$-7.802$                   &$3.281$            & $ -8.024$  \\
\bottomrule[1pt]
\end{tabular}}}
\end{center}
\end{table}

\subsection{Cosmological constant model}\label{sec4.1}
The cosmological constant $\Lambda$, corresponding to the energy density of quantum vacuum fluctuations, has become the leading explanation for dark energy since the discovery of cosmic acceleration in 1998. Despite facing some theoretical challenges \cite{Sahni:1999gb,Park:2017xbl}, it aligns well with various historical observations \cite{Aghanim:2018eyx}. The $\Lambda$CDM model, which incorporates $\Lambda$ and cold dark matter to describe the cosmic composition and expansion, is considered as the standard model of cosmology. The equation of state for $\Lambda$ is $-1$, and the reduced Hubble parameter for $\Lambda$CDM is given by
\begin{align}
E(z) = \left[ \Omega_{\text{m}}(1+z)^3 + (1 - \Omega_{\text{m}}) \right]^{1/2},
\end{align}
where $\Omega_{\rm m}$ is the matter density parameter. Note that the contribution from radiation is overlooked throughout the paper, because its influence becomes negligible compared to matter and dark energy in the late-time universe.

\begin{figure*}
\includegraphics[scale=0.9]{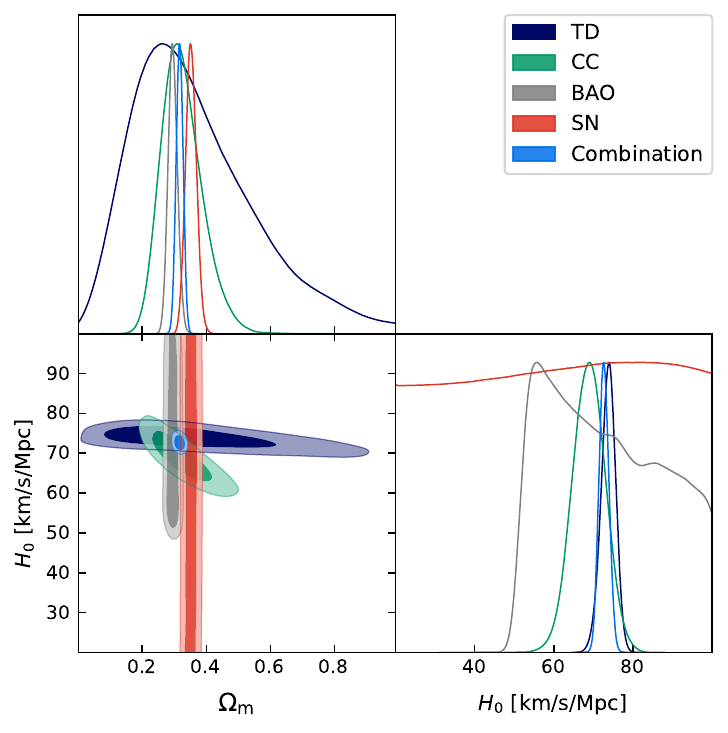}
\centering
\caption{The cosmological constant model: 68.3\% and 95.4\% confidence contours derived from individual TD, CC, BAO, and SN datasets, and their combined analysis (TD+CC+BAO+SN).}
\label{LCDM}
\end{figure*}

Using the observational data, we can obtain the best-fit values for cosmological parameters and the corresponding $\chi^2$ value as follows:
\begin{align}
\Omega_{\rm m}=0.317,\ H_0 = 72.7\,{\rm km/s/Mpc},\ \chi_{\min}^2 = 1686.282. \nonumber
\end{align}
We present the $1\sigma-2\sigma$ posterior distribution contours for the $\Lambda$CDM model in Fig.~\ref{LCDM}. As can be seen, the matter density parameter $\Omega_{\rm m}$ is consistent with the Planck CMB result, but the $H_0$ value is in around $4.1\sigma$ tension with that derived from the CMB observations \cite{Aghanim:2018eyx}. We note that measurement inconsistencies exist among late-universe observations themselves. For example, the BAO data provide $\Omega_{\rm m}=0.295^{+0.014}_{-0.015}$, while the SN data give $\Omega_{\rm m}=0.352\pm0.017$; there is a $2.5\sigma$ tension between the two results. While these values suggest a degree of inconsistency, it is below the commonly accepted 3$\sigma$ threshold for significant tension. Hence, it is considered appropriate to combine these datasets. Among all the models selected for analysis, $\Lambda$CDM is noted for having the highest $\chi_{\min}^2$. However, this does not indicate the poorest fit to the observational data. Generally, models with a greater number of parameters tend to have a lower $\chi_{\min}^2$ due to their increased flexibility in accommodating data. Considering its simplicity and sustained favorable performance in cosmological research, we choose the flat $\Lambda$CDM model as the reference for analyzing and comparing other models.

\subsection{Dark energy models with parameterized EoS}\label{sec4.2}
In this category, we explore three dark energy models with parameterized EoS: the constant $w$ ($w$CDM) model, the Chevallier--Polarski--Linder (CPL) model, and the Jassal--Bagla--Padmanabhan (JBP) model.
\subsubsection{Constant $w$ parametrization}\label{sec4.2.1}
In the $w$CDM model, the dark energy EoS is described by a constant $w$. The value of $w$ can be any real number, allowing for dark energy to have dynamical properties. For instance, if $w<-1$, the energy density of dark energy increase as the universe expands.
The growth in energy density engenders a repulsive force that not only continues to accelerate the cosmic expansion but does so at an ever-increasing rate. This scenario could potentially lead to the Big Rip, where the expansion rate is so rapid that it overcomes the gravitational forces that bind matter together, resulting in the disintegration of all structures in the universe. In contrast, when $w>-1$, the energy density of dark energy decreases as the universe expands. If $-1<w<-1/3$, the universe sustains a gradually weakening acceleration. For cases where $w\geq-1/3$, the universe reverts to the decelerated expansion, and in closed universe models, may ultimately lead to contraction in future epochs. The $w$CDM model introduces an additional degree of freedom compared to $\Lambda$CDM, which allows for more flexible fitting to the observational data. For this model, the reduced Hubble parameter is given by
\begin{align}
E(z) = \left[ \Omega_{\text{m}}(1+z)^3 + (1 - \Omega_{\text{m}})(1+z)^{3(1+w)} \right]^{1/2}.
\end{align}

\begin{figure*}
\includegraphics[scale=0.9]{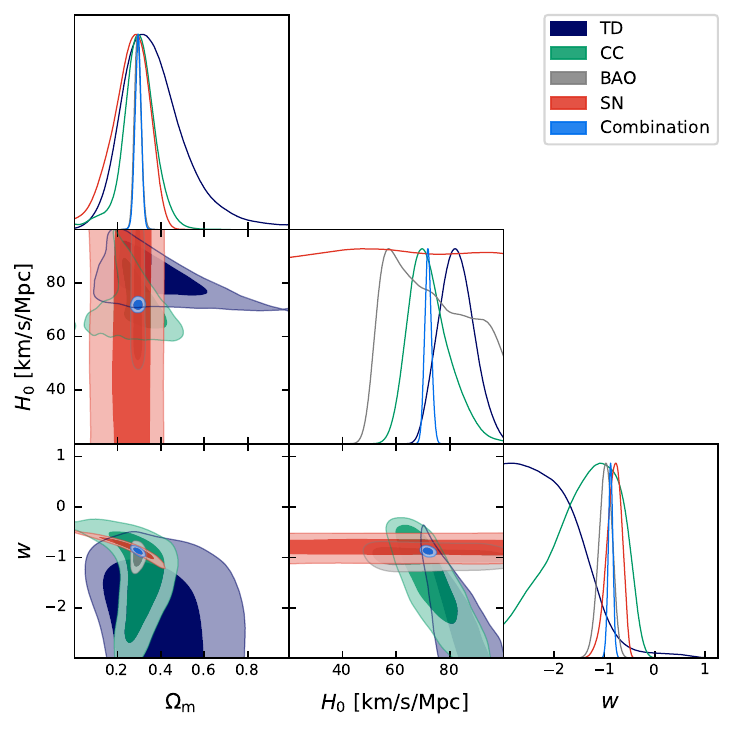}
\centering
\caption{The constant $w$ model: 68.3\% and 95.4\% confidence contours derived from individual TD, CC, BAO, and SN datasets, and their combined analysis (TD+CC+BAO+SN).}
\label{wCDM}
\end{figure*}

Using the observational data, we can get the best-fit parameters and the corresponding $\chi^2$ as follows:
\begin{align}
&\Omega_{\text{m}} = 0.295,\ H_0 = 71.9\,{\rm km/s/Mpc},\ w=-0.877, \nonumber \\
&\chi_{\min}^2 = 1679.190. \nonumber
\end{align}
The $1\sigma-2\sigma$ posterior distribution contours for $w$CDM are shown in Fig.~\ref{wCDM}. As can be seen, the central value of $w$ is situated within the quintessence-like regime, where $w>-1$. Furthermore, the cosmological constant scenario is excluded at approximately $2\sigma$ confidence level. Our results are consistent with the constraints reported in Refs.~\cite{DESI:2024mwx,DESI:2025fii,DES:2024tys}. Comprehensive analyses of $w$CDM using earlier observational data can be found in Refs.~\cite{Escamilla:2023oce,deCruzPerez:2024shj}. Compared to the flat $\Lambda$CDM model, the $w$CDM model yields $\Delta \rm AIC=-5.092$, $\Delta \rm DIC=-4.972$ and $\Delta \rm BIC=0.449$. Negative $\Delta{\rm AIC}$ and $\Delta{\rm DIC}$ values indicate that the model provides an improved fit to the observational data relative to the flat $\Lambda$CDM, with values between $-6$ and $-2$ constituting positive evidence for this improvement. A positive $\Delta \rm BIC$ value suggests an inferior fit to current data relative to the flat $\Lambda$CDM. However, since $\Delta \rm BIC<2$, the evidence against the $w$CDM model is only weak.

\subsubsection{The Chevallier--Polarski--Linder parametrization}\label{sec4.2.3}
The $w$CDM model serves as a foundational construct for exploring the dynamics of dark energy. However, the assumption of a constant EoS may not fully capture the complexity of dark energy's behavior \cite{Giare:2024gpk,Wolf:2025jlc,Wolf:2025jed,Paliathanasis:2025cuc,Paliathanasis:2025dcr}. This has motivated the introduction of more intricate models such as the CPL model \cite{Chevallier:2000qy,Linder:2002et}. It parameterizes the dark energy EoS as a linear function of the scale factor $a$, given by $w(z)=w_0+w_a(1-a)$, where $w_0$ and $w_a$ are free parameters. For this model, we have
\begin{align}
&E(z)= \left[\Omega_{\text{m}}(1+z)^3 \right. \nonumber \\
&\left. + (1 - \Omega_{\text{m}})(1+z)^{3(1+w_0+w_a)} \exp\left(-\frac{3w_a z}{1+z}\right)\right]^{1/2}.
\end{align}

\begin{figure*}
\includegraphics[scale=1]{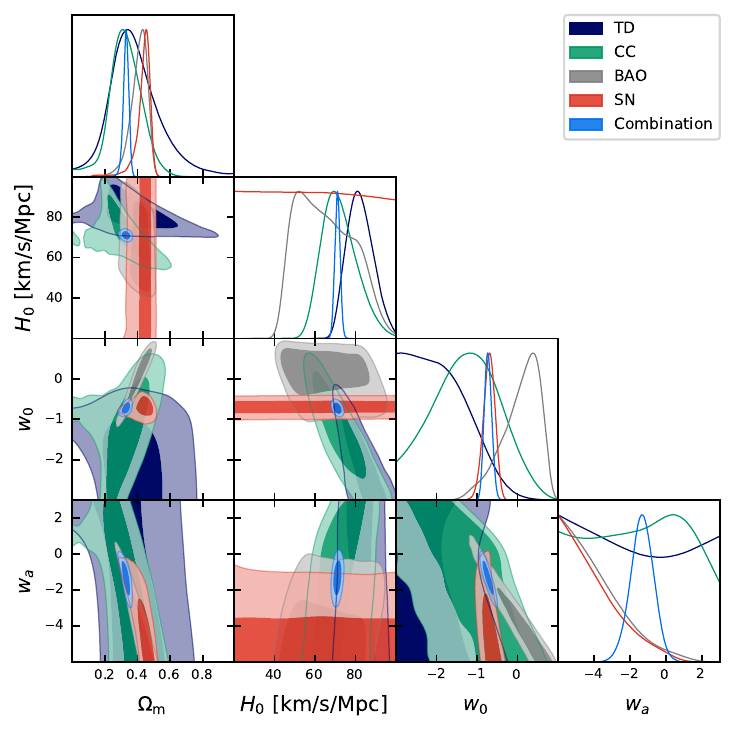}
\centering
\caption{The Chevallier--Polarski--Linder model: 68.3\% and 95.4\% confidence contours derived from individual TD, CC, BAO, and SN datasets, and their combined analysis (TD+CC+BAO+SN).}
\label{CPL}
\end{figure*}

Through the joint observational data analysis, we can obtain the best-fit parameters and the corresponding $\chi^2$ as:
\begin{align}
&\Omega_{\text{m}} = 0.329,\ H_0 = 71.0\,{\rm km/s/Mpc},\ w_0=-0.725, \nonumber \\
& w_a=-1.32,\ \chi_{\min}^2 = 1674.821. \nonumber
\end{align}
The $1\sigma-2\sigma$ likelihood contours for the CPL model are shown in Fig.~\ref{CPL}. As can be seen, the constraint results are not consistent with the $\Lambda$CDM model, i.e., the point ($w_0=-1$ and $w_a=0$) does not fall within the $1\sigma$ region. The cosmological constant is excluded at almost $3\sigma$ confidence level. The joint data favor the region with $w_0>-1$ and $w_a<0$, implying that EoS was phantom-like in the distant past and has evolved to quintessence-like at present. Our results are consistent with the recent studies, see Refs.~\cite{DESI:2024mwx,DESI:2025fii,DES:2024tys}. Note that pre-DESI analyses likewise indicated a statistical preference for dynamical dark energy \cite{Park:2024vrw,Park:2024pew}. The CPL model yields a $\Delta \rm AIC$ value of $-7.461$ and a $\Delta \rm DIC$ value of $-7.389$ when compared to the flat $\Lambda$CDM, providing strong evidence for an improved fit to the data. Both AIC and DIC penalize the CPL model for the number of model parameters, whereas BIC further penalizes the model for both the number of parameters and the sample size, leading to $\Delta \rm BIC=3.622$. The BIC result suggests that the observational data provide a support for the flat $\Lambda$CDM model over the CPL model, with $2<\Delta \rm BIC<6$ constituting positive evidence. The discrepancies in the outcomes of AIC/DIC and BIC primarily stem from their distinct evaluative strategies.

\subsubsection{The Jassal-Bagla-Padmanabhan parametrization}\label{sec4.2.3}
The JBP model was proposed to solve the high-$z$ issues inherent in the CPL model \cite{Jassal:2004ej}. It parameterizes the dark energy EoS as $w(z)=w_0+w_a z/(1+z)^2$, where $w_0$ and $w_a$ are also free parameters. In this model, the reduced Hubble parameter is calculated by
\begin{align}
&E(z) = \left[ \Omega_{\rm m}(1+z)^{3} \right. \nonumber \\
&+ \left. (1-\Omega_{\rm m})(1+z)^{3(1+w_{0})}\exp\left(\frac{3 w_{a} z^{2}}{2(1+z)^{2}}\right) \right]^{1/2}.
\end{align}

\begin{figure*}
\includegraphics[scale=1]{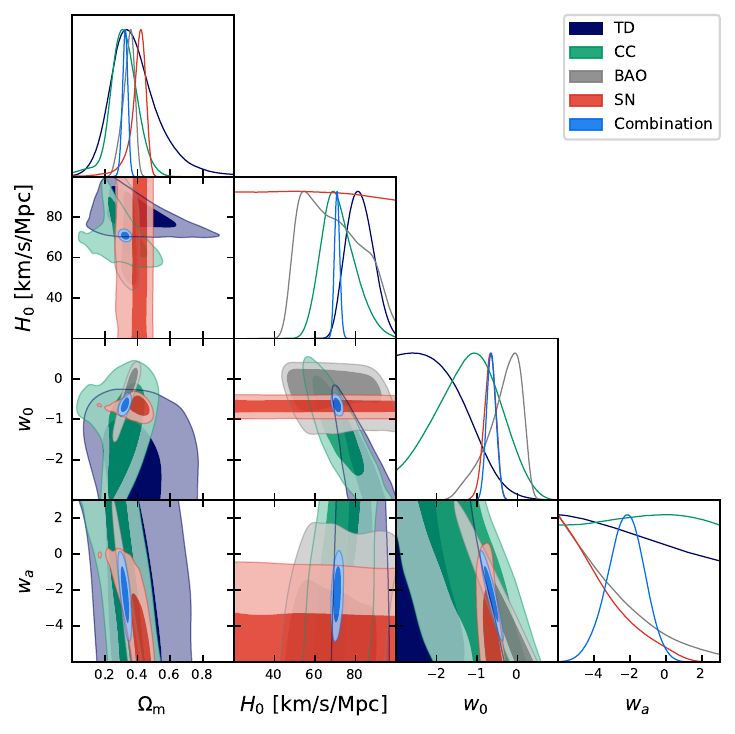}
\centering
\caption{The Jassal--Bagla--Padmanabhan model: 68.3\% and 95.4\% confidence contours derived from individual TD, CC, BAO, and SN datasets, and their combined analysis (TD+CC+BAO+SN).}
\label{JBP}
\end{figure*}

Based on the observational data, the best-fit parameters and the corresponding $\chi^2$ are:
\begin{align}
&\Omega_{\text{m}} = 0.324,\ H_0 = 70.9\,{\rm km/s/Mpc},\ w_0=-0.64, \nonumber \\
& w_a=-2.23,\ \chi_{\min}^2 = 1674.480. \nonumber
\end{align}
The $1\sigma-2\sigma$ likelihood contours for the JBP model are shown in Fig.~\ref{JBP}. We find that the constraint results are still inconsistent with the $\Lambda$CDM model. The best-fit EoS parameters for this model exhibit a greater deviation from the cosmological constant scenario than those of the CPL model. The combined analysis favors the parameter space with $w_0 > -1$ and $w_a < 0$. This solution exhibits tension with the previous study based on earlier BAO, SN and quasars~\cite{Bargiacchi:2021hdp}, which reported a preference for $w_0 < -1$ and $w_a > 0$.  Moreover, Ref.~\cite{Staicova:2022zuh} reports $w_0=-1.02\pm0.04$ and $w_a=0.22\pm0.23$ using the combination of early- and late-universe observations, consistent with the $\Lambda$CDM model. The JBP model yields the most negative $\Delta \rm AIC$ ($-7.802$) and $\Delta \rm DIC$ ($-8.024$) values, indicating that it is most favored by the observational data (with strong evidence). However, from the perspective of BIC, the difference between this model and $\Lambda$CDM reaches $3.281$, which implies that the current data favor the $\Lambda$CDM model over JBP (with positive evidence). The dynamical dark energy models, CPL and JBP, show similar performance, both strongly supported by AIC and DIC but significantly opposed by BIC.

\subsection{Dark energy models assuming a non-flat universe.}\label{sec4.3}
In this group, we explore two cosmological models: the $\Lambda$CDM model assuming a non-flat universe, denoted as o$\Lambda$CDM, and the $w$CDM model assuming a non-flat universe,  referred to as o$w$CDM.
\subsubsection{$\Lambda${\rm CDM} assuming a non-flat universe}\label{sec4.3.1}
Recent studies suggest that our universe may not be spatially flat. For instance, the CMB observations tend to favor a closed universe \cite{Aghanim:2018eyx,Handley:2019tkm,DiValentino:2019qzk}, while the late-time observations support an open universe \cite{Wu:2024faw}. Although these two measurements are inconsistent and even contradictory, they both support the concept of a non-Euclidean universe. A non-zero curvature would have far-reaching consequences for the inflationary paradigm and the ultimate fate of the universe. Therefore, incorporating the cosmic curvature into the parameter space is imperative when constraining cosmological parameters. For the o$\Lambda$CDM model, we have
\begin{align}
E(z) = \left[\Omega_{\text{m}}(1+z)^3 +\Omega_{K}(1+z)^2+ (1 - \Omega_{\text{m}}-\Omega_{K}) \right]^{1/2}.
\end{align}

\begin{figure*}
\includegraphics[scale=1]{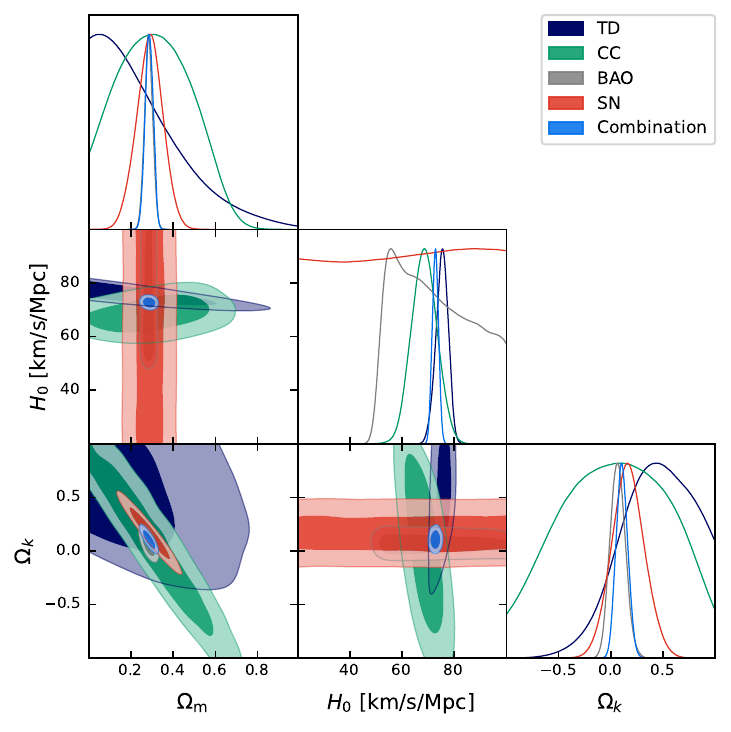}
\centering
\caption{The non-flat $\Lambda$CDM model: 68.3\% and 95.4\% confidence contours derived from individual TD, CC, BAO, and SN datasets, and their combined analysis (TD+CC+BAO+SN).}
\label{oLCDM}
\end{figure*}

By using the observational data, we can obtain the best-fit parameters and the corresponding $\chi^2$ as follows:
\begin{align}
&\Omega_{\text{m}} = 0.286,\ H_0 = 72.8\,{\rm km/s/Mpc},\,\Omega_K=0.106, \nonumber \\
&\chi_{\min}^2 = 1682.493. \nonumber
\end{align}
We present the $1\sigma-2\sigma$ posterior distribution contours for the o$\Lambda$CDM model in Fig.~\ref{oLCDM}. As shown in the $\Omega_{\rm m}-\Omega_{K}$ plane, the data combination favors an open universe, specially $\Omega_K=0.106\pm0.056$, which is in around $2.6\sigma$ tension with the Planck CMB result that our universe is closed, i.e., $\Omega_K=-0.044^{+0.018}_{-0.015}$. The two measurements have opposite signs, suggesting they are contrary to each other. Note that when the CC data is excluded, the derived constraints demonstrate increased statistical preference for an open universe, as highlighted in Ref.~\cite{Wu:2024faw}. It should be stressed that there are also some studies argue against the existence of a curvature tension, see Refs.~\cite{Efstathiou:2020wem,Liu:2024yib, deCruzPerez:2024shj,Fortunato:2025qxc}. Compared to the $\Lambda$CDM model, the o$\Lambda$CDM model yields $\Delta \rm AIC=-1.789$, $\Delta \rm DIC=-1.814$ and $\Delta \rm BIC=3.752$. Both the AIC and DIC analyses demonstrate a statistical preference for the non-flat $\Lambda$CDM model over the flat one (with weak evidence), while the BIC analysis provides positive evidence supporting the flat $\Lambda$CDM model as a more favorable model over the non-flat one.

\subsubsection{$w${\rm CDM} assuming a non-flat universe}\label{sec4.3.2}
It should be stressed that there is a strong degeneracy between the dark energy EoS and cosmic curvature. The current observational data leave room for deviations of the curvature parameter from zero (see Sec.~\ref{sec4.3.1}), which undermines the persuasiveness of the dark energy EoS constraints under the assumption of the flat universe scenario. Therefore, it is imperative to consider the spatial curvature parameter in the dynamic dark energy constraints. Here, we select the non-flat $w$CDM model to conduct our analysis. In this model, the reduced Hubble parameter can be written as
\begin{align}
E(z) &= \left[ \Omega_{\text{m}}(1+z)^3 +\Omega_{K}(1+z)^2 \right. \nonumber \\
&+ \left.(1 - \Omega_{\text{m}}-\Omega_{K})(1+z)^{3(1+w)} \right]^{1/2}.
\end{align}

\begin{figure*}
\includegraphics[scale=1]{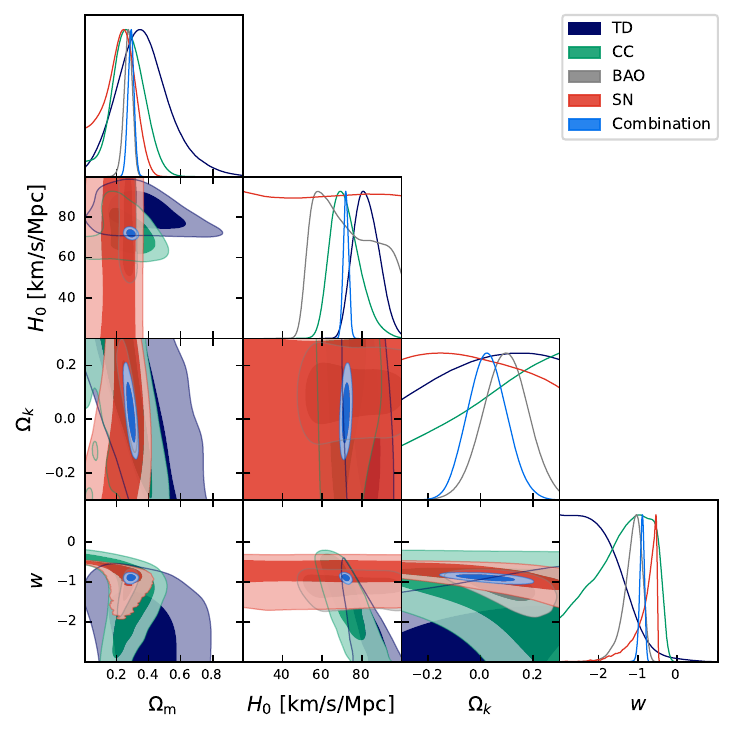}
\centering
\caption{The non-flat $w$CDM model: 68.3\% and 95.4\% confidence contours derived from individual TD, CC, BAO, and SN datasets, and their combined analysis (TD+CC+BAO+SN).}
\label{owCDM}
\end{figure*}

According to the sampling results, the optimal parameter estimates and the corresponding $\chi^2$ value are:
\begin{align}
&\Omega_{\text{m}} = 0.289,\ H_0 =  72.0\,{\rm km/s/Mpc},\, w=-0.890, \nonumber \\
& \Omega_K= 0.023, \,\chi_{\min}^2 = 1679.119. \nonumber
\end{align}
The $1\sigma-2\sigma$ posterior distribution contours for o$w$CDM are shown in Fig.~\ref{owCDM}. As can be seen, the constraint result is consistent with the spatially flat universe, specially $\Omega_K=0.023\pm0.073$. However, the central value of $\Omega_K$ is positive and has a significant deviation from zero. The flat universe can be ruled out at $>1\sigma$ confidence level even in the $w$CDM model by excluding the CC data, as discussed in Ref.~\cite{Wu:2024faw}. However, the authors pointed out that their constraints are dominated by the DESI BAO data, and the results require further verification with DESI's upcoming full-shape power spectrum. We also find that the central value of EoS parameter lies within the quintessence-like regime ($w>-1$), and the cosmological constant falls out the $2\sigma$ region. Our previous analysis revealed that the $w$CDM model is favored by AIC/DIC and not significantly opposed by BIC (see Sec.~\ref{sec4.2.1}). The o$w$CDM model has an additional parameter compared to $w$CDM and it yields a lower $\chi^2_{\rm min}$, but the difference is very small, specially $\Delta\chi^2_{\rm min}=-0.071$. The o$w$CDM model produces a $\Delta \rm AIC$ value of $-3.163$ and a $\Delta \rm DIC$ value of $-2.873$ when compared to the flat $\Lambda$CDM, suggesting that it provides a better fit to the observational data (with positive evidence). BIC further penalizes the model with sample size, leading to $\Delta \rm BIC=7.920$, which suggests that the current data provide a substantial support for the flat $\Lambda$CDM model over the o$w$CDM model, with a strong evidence ($\Delta \rm BIC>6$).

\subsection{Interacting dark energy models}\label{sec4.4}
The interacting dark energy models offer an extension to the standard model by introducing a dynamic interaction between dark energy and dark matter, potentially resolving theoretical cosmological issues and providing a more consistent picture with the observational data \cite{Valiviita:2008iv,Koyama:2009gd,Clemson:2011an,Li:2014eha, Li:2013bya,Xia:2016vnp,Giare:2024smz,Li:2024qso}. If there exists an interaction between dark sectors, the energy conservation equations can be written as
\begin{align}\label{IDEQ}
\dot{\rho}_{\rm de}&=-3H(1+w){\rho}_{\rm de}+Q, \nonumber \\
\dot{\rho}_{\rm c}&=-3H{\rho}_{\rm c}-Q,
\end{align}
where ${\rho}_{\rm de}$ and ${\rho}_{\rm c}$ are the energy densities of dark energy and dark matter, respectively, the dot denotes the derivative with respect to time, $w$ is the dark energy EoS, and $Q$ is the energy transfer rate. In this paper, we employ a phenomenological form of $Q=\beta H \rho_{\rm de}$, where $\beta$ is a dimensionless coupling parameter; $\beta>0$ indicates that dark matter decays into dark energy, $\beta<0$ denotes that dark energy decays into dark matter, and $\beta=0$ means that there is no interaction between them.

\subsubsection{$\Lambda${\rm CDM} with an interaction between dark sectors}\label{sec4.4.1}
In the context of the standard model, if we consider an interaction between dark energy and cold dark matter, the reduced Hubble parameter can be written as
\begin{align}
&E(z) = \left[\Omega_{\rm m}(1 + z)^3 \right. \nonumber \\
&+ \left.(1 - \Omega_{\rm m}) \left( \frac{\beta}{\beta+3} (1 + z)^3 + \frac{3}{\beta+3} (1 + z)^{-\beta} \right)\right]^{1/2}.
\end{align}
For simplicity, we refer to this model as I$\Lambda$CDM.

\begin{figure*}
\includegraphics[scale=0.9]{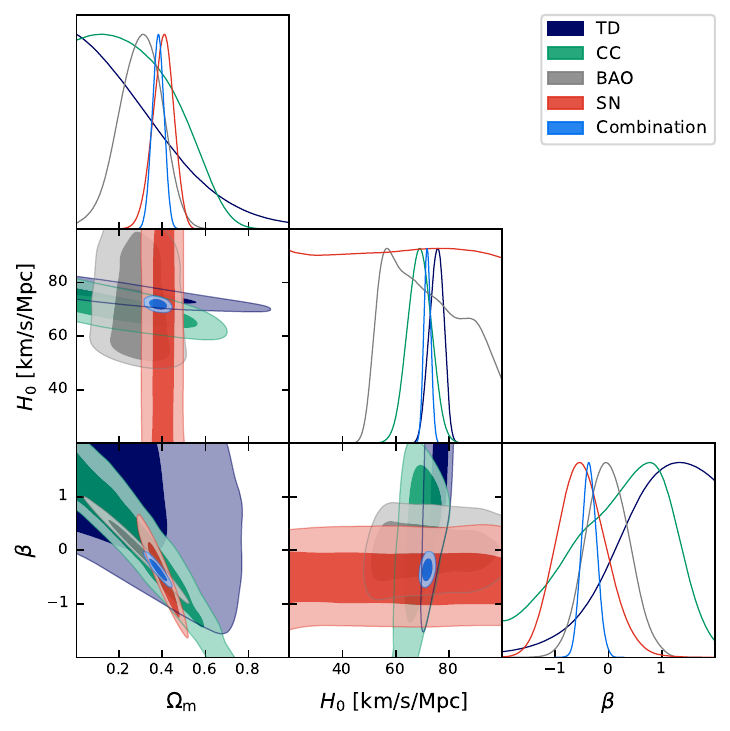}
\centering
\caption{The $\Lambda$CDM model with an interaction between dark energy and dark matter: 68.3\% and 95.4\% confidence contours derived from individual TD, CC, BAO, and SN datasets, and their combined analysis (TD+CC+BAO+SN).}
\label{ILCDM}
\end{figure*}

By using the joint data, we can obtain the best-fit cosmological parameters and the corresponding $\chi^2$ as:
\begin{align}
&\Omega_{\text{m}} = 0.380,\ H_0 =71.9\,{\rm km/s/Mpc},\,\beta=-0.36, \nonumber \\
&\chi_{\min}^2 = 1679.213. \nonumber
\end{align}
We present the $1\sigma-2\sigma$ posterior distribution contours for the I$\Lambda$CDM model in Fig.~\ref{ILCDM}. It can be seen that the observational data support the existence of an interaction between dark sectors at a confidence level exceeding $2\sigma$, with the conversion of dark energy into dark matter. This finding aligns with several recent studies; see Refs.~\cite{Li:2024qso,Pan:2025qwy,Giare:2024smz,Silva:2025hxw} for details. It is crucial to emphasize that, given the unknown nature of dark matter and dark energy, the consideration of interactions rely on assumed forms. Consequently, the existence of such interactions is inherently contingent upon the specific form assumed~\cite{DiValentino:2019ffd,DiValentino:2019jae,Giare:2024smz,Li:2024qso,Silva:2025hxw,Pan:2025qwy}. For this model, we have $\Delta \rm AIC=-5.069$,  $\Delta \rm DIC=-4.710$, and $\Delta \rm BIC=0.472$. Negative $\Delta \rm AIC$ and $\Delta \rm DIC$ values indicate that the model provides a substantial improvement fit to the observations than $\Lambda$CDM (with positive evidence). A positive $\Delta \rm BIC$ suggests $\Lambda$CDM is preferred over I$\Lambda$CDM, but a $\Delta \rm BIC$ value less $2$ is considered weak evidence against I$\Lambda$CDM.

\subsubsection{$w${\rm CDM} with an interaction between dark sectors}\label{sec4.4.1}
Considering that dark energy may not be the cosmological constant, it becomes imperative to extend investigations to dynamical dark energy models. If we consider an interaction between dark matter and dark energy in the $w$CDM model, the dimensionless Hubble parameter can be written as
\begin{align}
&E(z) = \left[\Omega_{\rm m}(1 + z)^3 + \right. \nonumber \\
&\left.(1 - \Omega_{\rm m}) \left( \frac{\beta}{\beta-3w} (1 + z)^3 + \frac{3w}{3w-\beta} (1 + z)^{(3+3w-\beta)} \right)\right]^{1/2}.
\end{align}
For simplicity, we refer to this model as I$w$CDM.

\begin{figure*}
\includegraphics[scale=1]{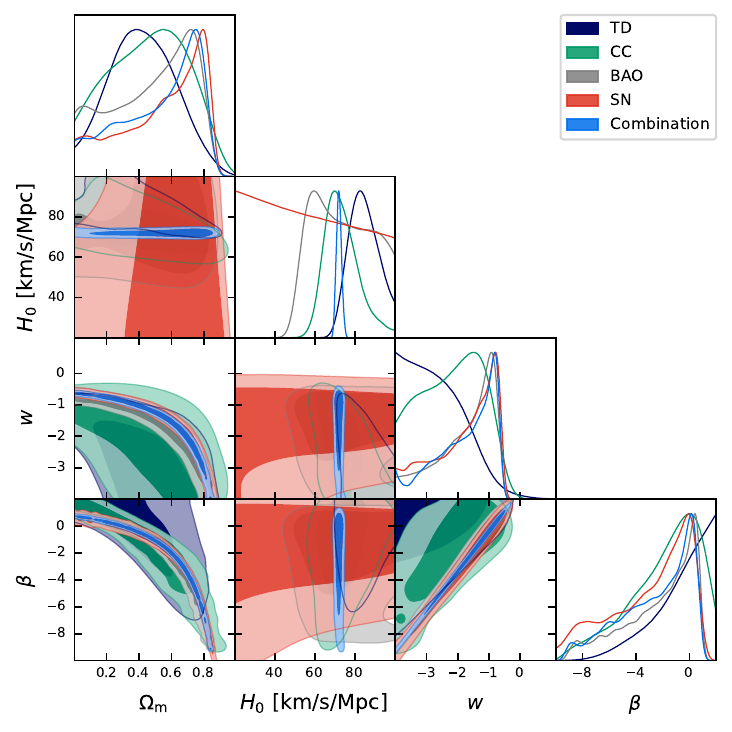}
\centering
\caption{The $w$CDM model with an interaction between dark energy and dark matter: 68.3\% and 95.4\% confidence contours derived from individual TD, CC, BAO, and SN datasets, and their combined analysis (TD+CC+BAO+SN).}
\label{IwCDM}
\end{figure*}

By using the observational data, the best-fit cosmological parameters and the corresponding $\chi^2$ are:
\begin{align}
&\Omega_{\text{m}} = 0.53,\ H_0 =71.9\,{\rm km/s/Mpc},\,w=-1.72\nonumber \\
&\beta=-2.5 ,\ \chi_{\min}^2 = 1679.181. \nonumber
\end{align}
The $1\sigma-2\sigma$ likelihood contours for the I$w$CDM model are shown in Fig.~\ref{IwCDM}. As can be seen, there is a strong degeneracy between the dark energy EoS parameter $w$ and the coupling parameter $\beta$, which impedes robust constraints on both of them. The constraint result is consistent with the $\Lambda$CDM model, i.e., the point ($w=-1$ and $\beta=0$) falls within the $1\sigma$ region. Interestingly, by allowing for interaction, the I$w$CDM model shifts the central value of dark energy EoS from quintessence (see Sec.~\ref{sec4.2.1}) to phantom. The I$w$CDM model has an additional parameter compared to the $w$CDM model and it yields a lower $\chi^2_{\rm min}$, but the difference is very small $\Delta\chi^2_{\rm min}=-0.009$. The I$w$CDM model produces a $\Delta \rm AIC$ value of $-3.101$ and a $\Delta \rm DIC$ value of $-4.644$  when compared to $\Lambda$CDM, indicating positive evidence for an improved fit to the observational data. BIC further penalizes the model with sample size, yielding $\Delta \rm BIC=7.920$, which indicates strong evidence against the I$w$CDM model ($\Delta \rm BIC>6$).

\subsection{Holographic dark energy models}\label{sec4.5}
In quantum field theory, the calculated vacuum energy density diverges. Even with the implementation of a reasonable ultraviolet (UV) cutoff, the calculated vacuum energy density remains approximately $120$ orders of magnitude greater than the observed value. This discrepancy arises from the absence of a complete and coherent theory of quantum gravity. The HDE model was proposed in this context, which integrates gravitational effects into the effective quantum field theory via the holographic principle \cite{Cohen:1998zx,Li:2004rb}. This principle posits that the number of degrees of freedom in a spatial region is finite to prevent black hole formation, which would otherwise occur with an excessive number of degrees. In the HDE model, the density of dark energy is determined by
\begin{align}
\rho_{\text{de}} \propto M_{\text{pl}}^2 L^{-2},
\end{align}
where $M_{\text{pl}}$ is the reduced Planck mass and $L$ is the infrared (IR) cutoff length scale in the effective quantum field theory. Consequently, the UV divergence issue encountered in the computation of vacuum energy density is effectively recast as an IR issue. The implementation of different IR cutoff prescriptions would result in varied cosmological consequences. This study employs two widely adopted computational methodologies for the IR cutoff.

\subsubsection{Holographic dark energy model}\label{sec4.5.1}
\begin{figure*}
\includegraphics[scale=0.9]{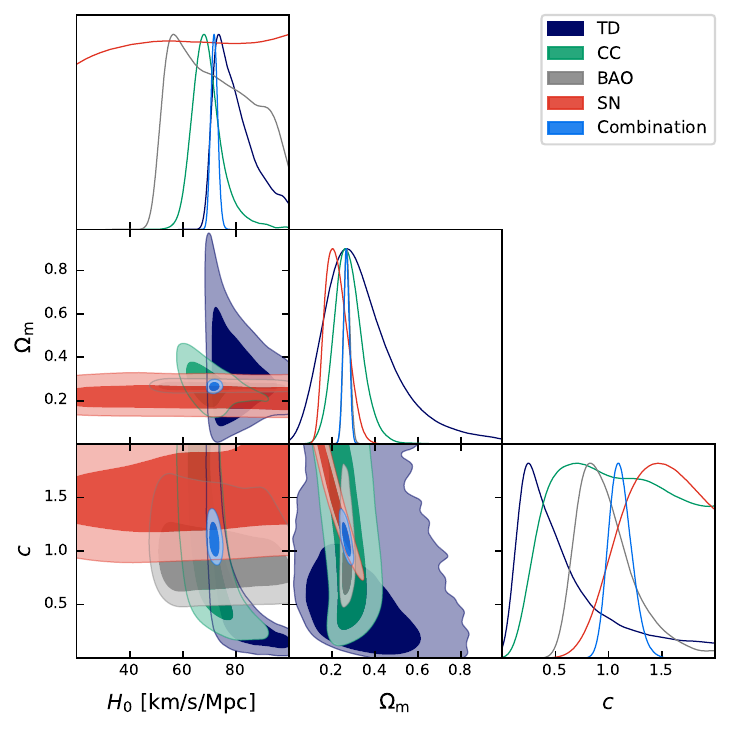}
\centering
\caption{The holographic dark energy model: 68.3\% and 95.4\% confidence contours derived from individual TD, CC, BAO, and SN datasets, and their combined analysis (TD+CC+BAO+SN).}
\label{HDE}
\end{figure*}

\begin{figure*}
\includegraphics[scale=0.9]{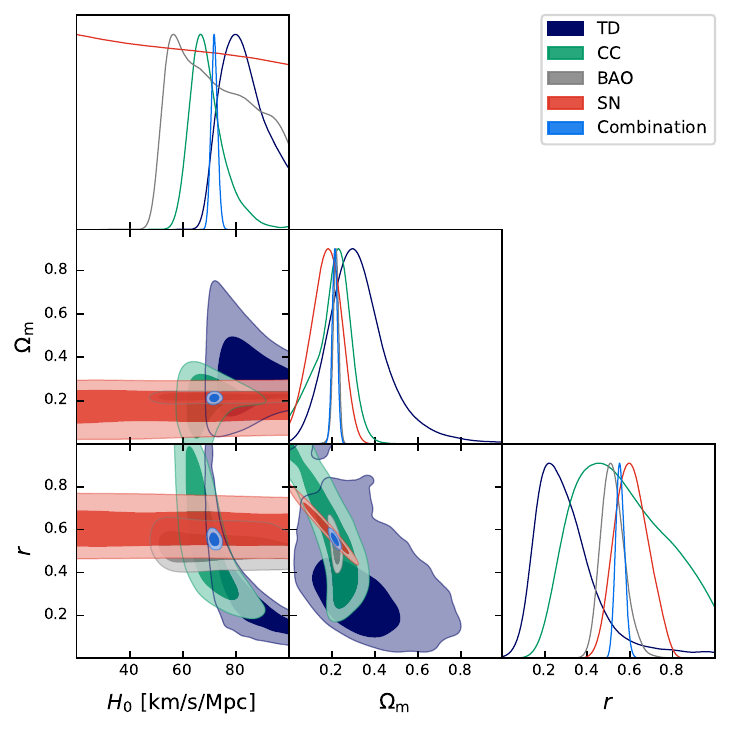}
\centering
\caption{The Ricci dark energy model: 68.3\% and 95.4\% confidence contours derived from individual TD, CC, BAO, and SN datasets, and their combined analysis (TD+CC+BAO+SN).}
\label{RDE}
\end{figure*}

The HDE model is formulated by employing the cosmic event horizon radius as the IR cutoff within holographic principle framework \cite{Li:2004rb}. In this model, the density of dark energy is given by
\begin{align}
\rho_{\text{de}} = 3c^2 M_{\text{pl}}^2 R_{\text{h}}^{-2},
\end{align}
where $c$ is a dimensionless parameter that influences the characteristics of HDE and $R_{\text{h}}$ is the future event horizon, calculated by
\begin{align}
R_{\text{h}}(a) = a \int_{a}^{\infty} \frac{{\rm d}a'}{H(a') a'^2}.
\end{align}
The dynamical evolution of dark energy is governed by the following differential equations
\begin{align}
\frac{1}{E(z)} \frac{{\rm d}E(z)}{{\rm d}z} &= -\frac{\Omega_{\text{de}}(z)}{1+z} \left( \frac{1}{2} + \frac{\sqrt{\Omega_{\text{de}}(z)}}{c} - \frac{3}{2\Omega_{\text{de}}(z)} \right) , \\
\frac{{\rm d}\Omega_{\text{de}}(z)}{{\rm d}z} = -&\frac{2\Omega_{\text{de}}(z)(1 - \Omega_{\text{de}}(z))}{1 + z} \times \left( \frac{1}{2} + \frac{\sqrt{\Omega_{\text{de}}(z)}}{c}  \right),
\end{align}
where $\Omega_{\text{de}}(z)$ is the fractional density of dark energy.

From the joint data analysis, we can get the best-fit parameters and the corresponding $\chi^2$ as:
\begin{align}
&\Omega_{\text{m}} = 0.266,\ H_0 =71.8\,{\rm km/s/Mpc},\,c=1.112\nonumber \\
& \chi_{\min}^2 = 1680.787. \nonumber
\end{align}
We plot the likelihood contours for the HDE model in Fig.~\ref{HDE}. Unlike the $\Lambda$CDM model, the EoS for the HDE model is dynamically evolving. The parameter $c$ is pivotal in determining the behavior of the dark energy EoS. When $c>1$, the EoS of HDE remains above $-1$, with HDE exhibiting quintessence-like behavior.  In contrast, for $c<1$, the EoS of HDE may cross the phantom divide at $w=-1$ in the future, leading to a universe dominated by phantom-like dark energy, which is predicted to end in a catastrophic Big Rip. As can be seen, the data combination shows a preference for $c>1$, indicating a quintessence-like behavior for HDE, and the universe is expected to undergo a relatively stable evolution. Our findings align with the latest HDE research \cite{Li:2024qus,Li:2024hrv,Tang:2024gtq}. Nevertheless, several critical discussions are warranted: Current late-universe observations favor $c>1$, whereas early-universe observations support $c < 1$. This apparent inconsistency suggests two possible interpretations --- either the HDE framework fails to accurately describe the cosmological evolution, or undetected systematic errors exist in measurements. We plan to conduct an investigation of this issue in future works. The HDE model yields $\Delta \rm AIC= -3.495$, $\Delta \rm DIC= -3.399$ and $\Delta \rm BIC=2.046$. As can be seen, while AIC and DIC metrics provide positive evidence favoring the HDE model, BIC provides comparable evidence against HDE due to its heavier penalty on the model complexity.

\subsubsection{Ricci dark energy model}\label{sec4.5.2}
\begin{figure*}
\includegraphics[scale=0.9]{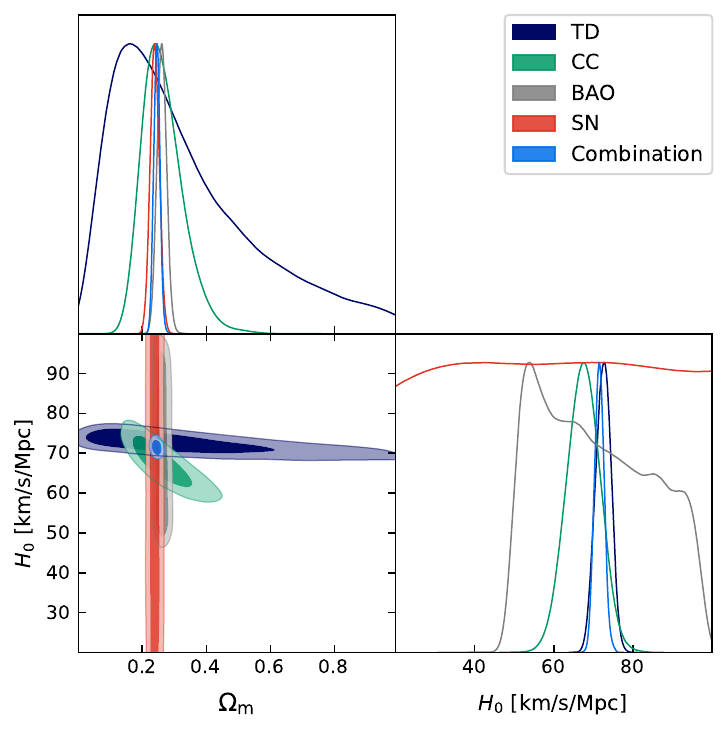}
\centering
\caption{The Dvali--Gabadadze--Porrati model: 68.3\% and 95.4\% confidence contours derived from individual TD, CC, BAO, and SN datasets, and their combined analysis (TD+CC+BAO+SN).}
\label{DGP}
\end{figure*}
The RDE model proposes that the dark energy density is proportional to the Ricci scalar, which is a measure of the curvature of spacetime \cite{Gao:2007ep}. Unlike the HDE model, which is based on the entropy bound of quantum systems, RDE is rooted in the geometry of spacetime itself. This makes RDE an attractive model for further exploration, as it may offer insights into the interplay between quantum mechanics and general relativity, which is crucial for understanding the universe at its fundamental level. The RDE model chooses the mean radius of the Ricci scalar curvature as the IR cutoff length scale, and the dark energy density is given by
\begin{align}
\rho_{\rm de} = 3\gamma M_{\rm pl}^2 (\dot{H} + 2H^2),
\end{align}
where $\gamma$ is a positive constant. The evolution of the Hubble parameter is determined by the following differential equation,
\begin{align}
E^{2} = \Omega_{\rm m} e^{-3x} + \gamma \left( \frac{1}{2} \frac{{\rm d}E^{2}}{{\rm d}x} + 2E^{2} \right),
\end{align}
where $x\equiv\ln a$. Solving this equation, we obtain
\begin{align}
E(z) = \left( \frac{2\Omega_{\rm m}}{2-\gamma}(1+z)^{3} + \left(1-\frac{2\Omega_{\rm m}}{2-\gamma}\right)(1+z)^{\left(4-\frac{2}{\gamma}\right)} \right)^{1/2}.
\end{align}

Using the observational data, we can derive the best-fit values for cosmological parameters and the corresponding $\chi^2$ as follows:
\begin{align}
&\Omega_{\text{m}} = 0.212,\ H_0 =71.8\,{\rm km/s/Mpc},\,\gamma=0.554\nonumber \\
& \chi_{\min}^2 = 1679.183. \nonumber
\end{align}
The $1\sigma-2\sigma$ likelihood contours for the RDE model are shown in Fig.~\ref{RDE}. The RDE model is capable of avoiding the causality problem faced by the HDE model, and it can also naturally resolve the coincidence problem. The parameter $\gamma$ plays a crucial role in determining the behavior of dark energy. For example, during the dark energy-dominated epoch, $\gamma>0.5$ leads to $w>-1$, while $\gamma<0.5$ would result in $w<-1$. As can be seen, the data combination shows a preference for $\gamma>0.5$, indicating a quintessence-like behavior for RDE. Ref.~\cite{Li:2024hrv} establishes joint early- and late-universe constraints on the RDE model, with strong evidence indicating $\gamma<0.5$. This discrepancy may stem from early-time observations tending to favor $\gamma< 0.5$. A previous study, utilizing the observational data available at the time (inclusive of the CMB data), robustly ruled out the RDE model \cite{Xu:2016grp}. However, the data we utilized here provide a contrary conclusion. Compared to the $\Lambda$CDM model, the RDE model yields $\Delta \rm AIC= -5.092$, $\Delta \rm DIC= -5.069$ and $\Delta \rm BIC=0.442$. Negative $\Delta \rm AIC$ and $\Delta \rm DIC$ values indicate that the RDE model offers a better fit to the current data (with positive evidence) compared to the $\Lambda$CDM model. A positive $\Delta \rm BIC$  value suggests a preference for the $\Lambda$CDM model over the RDE model, but only with weak evidence ($\Delta \rm BIC<2$). Therefore, the RDE model demonstrates commendable performance and warrants further exploration and attention.

\subsection{Effective dark energy models (MG theories)}\label{sec4.6}

In physics, the observed acceleration of the universe is commonly attributed to one of two possibilities: the presence of dark energy or the deviations from Einstein's general relativity on cosmological scales. The latter, referred to as MG theories, can create effective dark energy scenarios mimicking the behavior of actual dark energy. As a crucial avenue for explaining the acceleration, it is imperative to extend our investigation into MG models. A prominent exemplar of such theoretical constructs is the DGP model \cite{Dvali:2000hr}. It posits that gravity can leak from our four-dimensional Minkowski brane into an extra dimension of a five-dimensional bulk space-time. In this subsection, we constrain the DGP model and one of its phenomenological extensions.

\begin{figure*}
\includegraphics[scale=0.9]{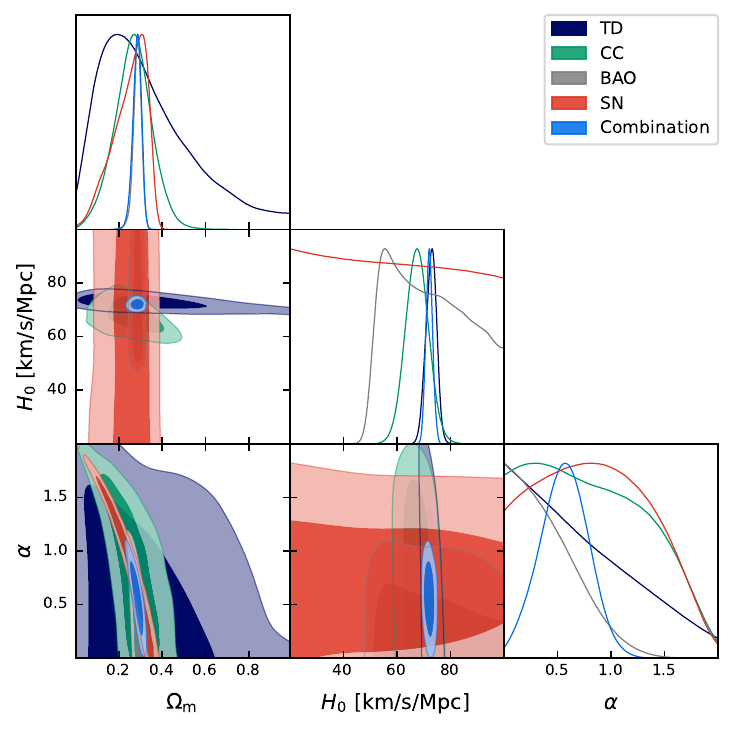}
\centering
\caption{The $\alpha$DE model: 68.3\% and 95.4\% confidence contours derived from individual TD, CC, BAO, and SN datasets, and their combined analysis (TD+CC+BAO+SN).}
\label{alphaDE}
\end{figure*}

\subsubsection{Dvali-Gabadadze-Porrati model}\label{sec4.6.1}

In the DGP model, the Friedmann equation is modified as
\begin{align}
3 M_{\rm pl}^{2}\left(H^{2}-\frac{H}{r_{\rm c}}\right)=\rho_{\rm m}(1+z)^{3},
\end{align}
where $r_{\rm c} = \left[ H_{0} (1 - \Omega_{\rm m}) \right]^{-1}$ is the length scale at which the phenomenon of gravitational leakage into the extra dimension occurs. Thus, the reduced Hubble parameter can be written as
\begin{align}
E(z) = \sqrt{\Omega_{\text{m}}(1+z)^3 + \Omega_{r_{\rm c}}} + \sqrt{\Omega_{r_{\rm c}}},
\end{align}
where $\Omega_{r_{\rm c}} = (1 - \Omega_{\rm m})^2/4$.

From the joint constraints, we can get the best-fit parameters and the corresponding $\chi^2$:
\begin{align}
&\Omega_{\text{m}} = 0.245,\ H_0 = 71.5\,{\rm km/s/Mpc},\, \chi_{\min}^2 = 1684.453. \nonumber
\end{align}
We present the $1\sigma-2\sigma$ posterior distribution contours for the DGP model in Fig.~\ref{DGP}. As can be seen, the results tend toward a relatively low value of $\Omega_{\rm m}$, which is consistent with the findings reported in Ref.~\cite{DES:2024fdw}. The DGP model shares an equivalent number of parameters with the flat $\Lambda$CDM model. Compared to flat $\Lambda$CDM, the DGP model yields a smaller $\chi^2_{\rm min}$, indicating a superior fit to the observational data. This model is the only one that supported by both the AIC, BIC and DIC metrics, with $\Delta \rm AIC= -1.829$, $\Delta \rm BIC= -1.829$ and $\Delta \rm DIC= -1.798$, though the magnitude of these values falls just below the threshold for positive evidence. It should be noted that this model is strongly disfavored when the CMB data are incorporated, as found in Refs.~\cite{Xu:2016grp,Zhai:2017vvt,Li:2011yy,Lombriser:2009xg}.

\subsubsection{$\alpha$ dark energy model}\label{sec4.6.2}
The $\alpha$DE model emerges as a phenomenological extension of the DGP model, in which the Friedmann equation is modified as
\begin{align}
3 M_{\text{\rm pl}}^2 \left( H^2 - \frac{H^\alpha}{r_{\rm c}^{2-\alpha}} \right) = \rho_{\rm m} (1 + z)^3,
\end{align}
where $\alpha$ is a phenomenological parameter and $r_{\rm c} = (1 - \Omega_{\rm m})^{1/(\alpha-2)} H_0^{-1}$. The reduced Hubble parameter can be derived from the equation
\begin{align}
E(z)^2 = \Omega_{\rm m} (1 + z)^3 + E(z)^\alpha (1 - \Omega_{\rm m}).
\end{align}
It is evident that the $\alpha$DE model transitions to the DGP model when the parameter $\alpha$ is set to $1$, and it reduces to the $\Lambda$CDM model when $\alpha$ is taken as $0$.

From the joint constraints, we obtain the best-fit parameters and the corresponding $\chi^2$:
\begin{align}
&\Omega_{\text{m}} = 0.284,\ H_0 = 72.1\,{\rm km/s/Mpc},\,\alpha=0.56, \nonumber \\
&\chi_{\min}^2 = 1680.915. \nonumber
\end{align}
We present the $1\sigma-2\sigma$ posterior distribution contours for the $\alpha$DE model in Fig.~\ref{alphaDE}. As can be seen, the DGP ($\alpha=1$) and $\Lambda$CDM ($\alpha=0$) models are both ruled out by the current observational data at about $2\sigma$ confidence level. Compared to flat $\Lambda$CDM, the $\alpha$DE model produces $\Delta \rm AIC=-3.367$, $\Delta \rm DIC=-3.547$ and $\Delta \rm BIC= 2.175$. The AIC and DIC analyses indicate a preference for the $\alpha$DE model (with positive evidence), whereas $\Delta \rm BIC>2$ suggests positive evidence against the $\alpha$DE model.

\section{Discussion and conclusion}\label{sec5}

\begin{figure*}
\includegraphics[scale=0.8]{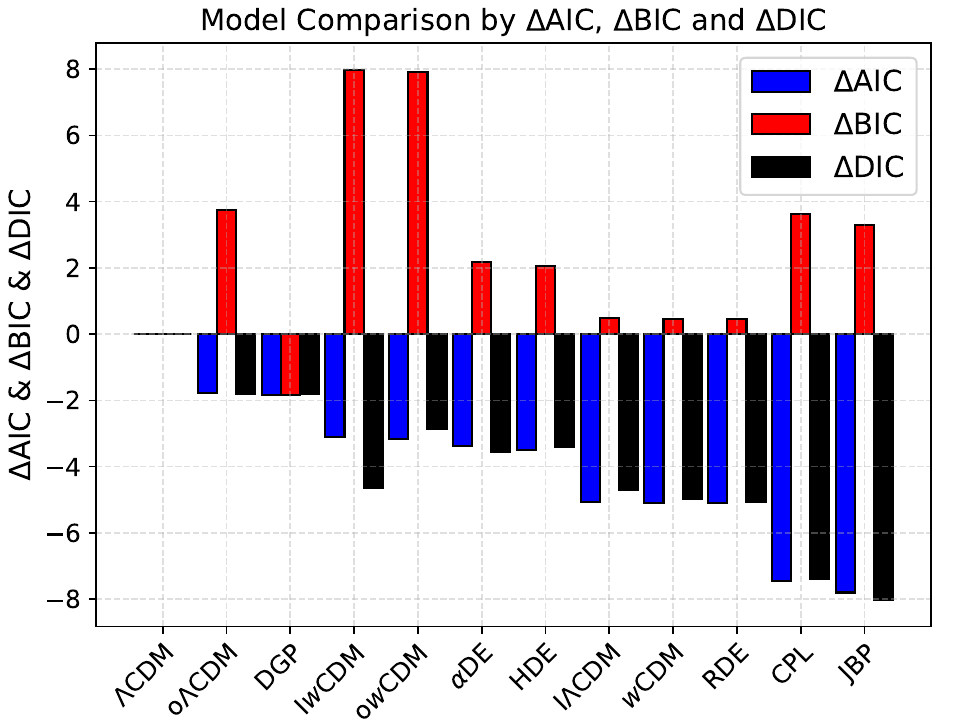}
\centering
\caption{Graphical representation of model comparison results. In this work, we choose flat $\Lambda$CDM as a reference model to calculate the $\Delta$AIC, $\Delta$BIC and $\Delta$DIC values. Generally, a negative $\Delta{\rm A/B/DIC}$ value within the range $[-2, 0)$ indicates weak evidence in favor of the model, a value within $[-6, -2)$ indicates positive evidence in favor, a value within $[-10, -6)$ indicates strong evidence in favor, and a value less than $-10$ indicates very strong evidence in favor. Conversely, a $\Delta{\rm A/B/DIC}$ value within the range $(0, 2]$ indicates weak evidence against the model, a value within $(2, 6]$ indicates positive evidence against, a value within $(6, 10]$ indicates strong evidence against, and a value greater than 10 indicates very strong evidence against. The order of models from left to right is arranged according to the values of $\Delta$AIC, i.e., in order of decreasing $\Delta$AIC.}
\label{AICBICDICfigures}
\end{figure*}

The $\Lambda$CDM model remains the leading framework for dark energy due to its robust performance and consistent agreement with various observations. However, recent observations suggest that some competing dark energy models could potentially challenge its dominance. In this paper, we explore whether any of these models can match or even surpass the $\Lambda$CDM model in terms of fitting the current observational data. Given the discrepancies observed in measurements between the early- and late-time universe, our analysis is conducted using only the late-universe observations, including the baryon acoustic oscillation, type Ia supernova, cosmic chronometer, and strong gravitational lensing time delay data. The paper presents a thorough evaluation of twelve prominent cosmological models.

The dark energy models vary in their parameter count. Compared to the $\Lambda$CDM model, although these models can yield lower $\chi^2$ values, this does not necessarily indicate superior performance. The reduced $\chi^2$ values could be a consequence of overfitting, where models have captured the noise in data rather than the underlying patterns. Therefore, we employ AIC, BIC and DIC to penalize the model complexity, thereby ensuring a fair assessment that balances goodness of fit with model simplicity. The model comparison results using the information criteria are summarized in Table~\ref{tab:results2}. To enhance visual comprehension, we present the numerical results in a graphical format in Fig.~\ref{AICBICDICfigures}. We are only concerned with the relative values of AIC, BIC and DIC, rather than their absolute values. We calculate the $\Delta$AIC, $\Delta$BIC and $\Delta$DIC values for each model, using the flat $\Lambda$CDM model as the reference.

From the perspective of AIC and DIC, all of the dark energy models perform better than the flat $\Lambda$CDM model. Generally, compared to the flat $\Lambda$CDM model, a model with $-2\leq\Delta{\rm AIC,DIC}<0$ is favored with weak evidence, $-6\leq\Delta{\rm AIC,DIC}<-2$ indicates positive evidence in favor, $-10\leq\Delta{\rm AIC,DIC}<-6$ indicates strong evidence in favor, and $\Delta{\rm AIC,DIC}<-10$ indicates very strong evidence in favor. All the dark energy models yield negative $\Delta \rm AIC$ and $\Delta \rm DIC$ values, indicating that although these models have been penalized by AIC and DIC for their increased complexity, they offer a better fit to the observational data. The dynamical dark energy models, specifically the CPL and JBP models, have demonstrated the best performance among all the cosmological models, and the constraint results indicate that the cosmological constant scenario is excluded at the $3\sigma$ confidence level. From the results, we find compelling evidence forcing us away from the cosmological constant.

From the standpoint of BIC, the flat $\Lambda$CDM continues to exhibit excellent performance, with only one model being more favored. Typically, a BIC difference of $2$ is generally indicative of significant evidence against the model with the higher BIC, while a difference within $(2, 6]$ indicates positive evidence against, a value within $(6, 10]$ indicates strong evidence against, and a value greater than 10 indicates very strong evidence against. Compared to the flat $\Lambda$CDM model, two models are strongly disfavored (I$w$CDM, o$w$CDM; $6<\Delta\rm BIC\leq10$), five models are significantly disfavored (CPL, JBP, o$\Lambda$CDM, $\alpha$DE, HDE; $2<\Delta\rm BIC\leq6$), three models are weakly disfavored (I$\Lambda$CDM, $w$CDM, RDE; $0<\Delta\rm BIC <1$), and one model outperforms the $\Lambda$CDM model (DGP; $\Delta\rm BIC=-1.829$). Overall, the BIC analysis leaves substantial room for the flat $\Lambda$CDM model.

Taking into account both the AIC, BIC and DIC analysis, only one model performs better than the flat $\Lambda$CDM model, i.e., DGP. The model outperforms all dark energy models, which implies that modified gravity theories, one of the two approaches to explaining the accelerated expansion of the universe, merit significant attention. In addition, the I$\Lambda$CDM, $w$CDM, and RDE models are robust contenders to the $\Lambda$CDM model, all receiving substantial support from AIC and DIC, and not being significantly disapproved by BIC. Beyond these, all other models are supported by AIC and DIC (from weak to strong evidence) but disfavored by BIC (from positive to strong evidence). The divergent model selection preferences shown by AIC, BIC and DIC can be attributed to the lighter penalty of AIC and DIC on the number of model parameters, which tends to favor more complex models, while BIC imposes a heavier penalty on the model complexity, thus favoring simpler models.

We have detected numerous indications of new physics beyond the standard model, such as the possibility that the universe may be spatially open, dark energy may be dynamically evolving, there may be an interaction between dark matter and dark energy, all at high confidence levels. Additionally, we have explored the nature of dark energy from the perspective of quantum gravity, and this framework is well-received by the current observational data, especially the RDE model. It should be noted that some models cannot be effectively constrained. For instance, when exploring the spatial flatness of the universe or the interaction between dark sectors within the $w$CDM framework, we are unable to provide tight constraints. In cosmological research, considering multiple effects simultaneously is important due to the parameter degeneracy among these effects; however, the current data does not support such an approach. We can only hope for more data or the emergence of new probes to enable this \cite{Wu:2022dgy}.

In physics, the cosmic acceleration can be attributed to two primary mechanisms: either the presence of dark energy or modifications to the gravity on cosmological scales. In the present work, we primarily study dark energy models, while also considering two MG models, which we refer to as effective dark energy models that can mimic the behavior of dark energy. Surprisingly, the results indicate that one of the MG models, the DGP model, outperforms all the dark energy models we consider. In the forthcoming research, we plan to investigate a broader range of MG theories, exploring diverse modifications to gravity on large scales, to discern which scenarios are more favored by the current observational data.

\begin{acknowledgments}
We thank Tian-Nuo Li and Guo-Hong Du for fruitful discussions.

\end{acknowledgments}

\bibliography{EoS}

\end{document}